\documentclass[pra,aps,amsmath,amssymb,amsfonts,twocolumn,nofootinbib,longbibliography,superscriptaddress]{revtex4-2}
\usepackage{amsfonts}
\usepackage{txfonts}
\usepackage{amsmath}
\usepackage{float}
\usepackage{diagbox}
\usepackage[colorlinks,breaklinks,linkcolor=blue,anchorcolor=blue,citecolor=blue,urlcolor=blue]{hyperref}
\usepackage{graphicx}
\usepackage{dcolumn}
\usepackage{bm}
\usepackage{amssymb}
\usepackage{mathrsfs}
\usepackage[sort&compress]{natbib}
\usepackage{subfigure}
\usepackage{braket}
\usepackage{xcolor}

\usepackage{lineno}
 
\begin{document}

\title{Nonreciprocal magnon-magnon entanglement in a spinning cavity-magnon system}
\author{Zhisheng Xu}
\affiliation{Center for Quantum Sciences and School of Physics, Northeast Normal University, Changchun 130024, China}

\author{Mengxue Li}
\affiliation{Center for Quantum Sciences and School of Physics, Northeast Normal University, Changchun 130024, China}

\author{Chunfang Sun}
\email{suncf997@nenu.edu.cn}
\affiliation{Center for Quantum Sciences and School of Physics, Northeast Normal University, Changchun 130024, China}

\author{Gangcheng Wang}
\email{wanggc887@nenu.edu.cn}
\affiliation{Center for Quantum Sciences and School of Physics, Northeast Normal University, Changchun 130024, China}

\date{\today}

\begin{abstract}
Cavity-magnon systems, combining magnons and photons, offer a versatile platform for studying quantum entanglement and advancing quantum information science. In this work, we propose a scheme for generating nonreciprocal magnon-magnon entanglement in a hybrid system consisting of two yttrium iron garnet spheres coupled to a spinning whispering-gallery-mode cavity. By leveraging the magnon Kerr nonlinearity and the Sagnac effect arising from the cavity rotation, we show that the entanglement can be substantially enhanced, and the resulting entanglement exhibits pronounced nonreciprocal characteristics. Furthermore, our scheme demonstrates that the entanglement remains robust against thermal noise and persists at bath temperatures up to 100 mK. This work underscores the potential of spinning cavity-magnon systems as a versatile platform for realizing nonreciprocal quantum devices and facilitating the development of quantum technologies.
\end{abstract}

\maketitle
\section{Introduction}
\label{SecI}
Quantum entanglement \cite{RevModPhys.81.865,vedral2014quantum}, one of the most profound and distinctive phenomena in quantum mechanics, has emerged as a cornerstone of modern quantum information science. As a vital resource, it underpins a broad spectrum of quantum technologies such as quantum computing \cite{steane1998quantum,rietsche2022quantum}, and quantum communication \cite{brassard2003quantum,gisin2007quantum}. In recent decades, substantial research efforts have been directed toward realizing and controlling entangled states across diverse physical platforms. In particular, macroscopic quantum entanglement \cite{lvovsky2013observation,PhysRevLett.114.057203,PhysRevA.94.053807} has garnered significant attention due to its promising applications in quantum transduction \cite{PhysRevApplied.18.054061,PhysRevLett.129.215901}, quantum networks \cite{kimble2008quantum}, quantum sensing \cite{lawrie2019quantum,Zhang2021,RevModPhys.89.035002}, Bell tests \cite{PhysRevLett.122.090402}, quantum teleportation \cite{PhysRevLett.70.1895}, and microwave-to-optical conversion \cite{PhysRevA.96.013808,eshaqi2022nonreciprocal}.

Magnons \cite{chumak2015magnon,YUAN20221,ZARERAMESHTI20221}, the quanta of collective spin excitations in magnetically ordered materials—particularly in yttrium iron garnet (YIG) \cite{mallmann2013,10.1063/1.5144202}—offer a promising platform to explore quantum phenomena. Owing to their high spin density and low dissipation, magnons in YIG spheres have enabled the investigation of diverse effects such as dissipative coupling \cite{10.1063/5.0046202}, dark modes \cite{zhang2015magnon}, the magnon Kerr effect \cite{PhysRevApplied.12.034001}, exceptional points \cite{doi:10.1126/science.aar7709,chen2017exceptional,PhysRevB.99.054404,PhysRevApplied.15.024056}, unidirectional quantum switching \cite{5357489}, and spin current manipulation \cite{doi:10.1098/rsta.2011.0010}. The magnon Kerr effect, originating from nonlinear magnon-magnon interactions due to magnetic anisotropy, exhibits strong nonlinear behavior that is essential for macroscopic quantum entanglement. Consequently, magnon systems have been extensively employed in cavity-magnon systems \cite{doi:10.1098/rsta.2011.0105,Flower2019}, cavity optomechanics \cite{PhysRevLett.105.133602,RevModPhys.86.1391,10.1063/1.4896029,PhysRevLett.121.203601,CHEN2026117472}, and cavity magnomechanics \cite{PhysRevA.99.021801} to investigate macroscopic quantum entanglement. Moreover, Kerr nonlinearity \cite{PhysRevLett.91.093601} has significantly advanced research on multistability \cite{PhysRevLett.129.123601}, long-range spin-spin interactions \cite{PhysRevA.107.033516}, quantum phase transitions \cite{PhysRevB.104.064423,PhysRevA.108.033704}, and high-sensitivity detection \cite{PhysRevB.107.064417}.

Meanwhile, nonreciprocal phenomena, which break Lorentz reciprocity by exhibiting distinct physical behaviors when the driving field propagates in opposite directions within the cavity, have found ever-wider applications in cavity-magnon systems, such as nonreciprocal entanglement \cite{PhysRevB.108.024105,PhysRevA.108.063704,PhysRevA.109.043512,PhysRevA.111.013713,PhysRevA.105.013711,PhysRevA.110.063711}, nonreciprocal phonon \cite{PhysRevA.103.053501} and nonreciprocal magnon lasers \cite{Huang:22}. Optical devices designed based on such nonreciprocal effects have become indispensable for various practical applications, including communication technologies \cite{PhysRevA.97.052315}, sensing \cite{Jing:18}, and optical signal processing \cite{PhysRevApplied.10.047001}.

Quantum entanglement in a spinning whispering-gallery-mode (WGM) cavity-magnon system \cite{https://doi.org/10.1002/adom.202100143,PhysRevB.96.094412} can be significantly enhanced and protected. In these systems, photon-magnon correlations become asymmetric with respect to the propagation direction, resulting in a phenomenon known as nonreciprocal entanglement. This directional dependence arises from the Sagnac–Fizeau effect \cite{RevModPhys.39.475}, which introduces opposite frequency shifts in the cavity mode. When the cavity rotation direction differs from that of the driving field, the effective refractive index experienced by the intracavity light becomes direction-dependent, breaking Lorentz reciprocity and resulting in nonreciprocal correlations.

In addition to the Sagnac effect, the magnon Kerr effect provides another mechanism for achieving nonreciprocal entanglement within the cavity. Over the past decade, nonreciprocal entanglement has been extensively studied in cavity-magnon systems, including magnon-photon, magnon-phonon, and photon-phonon entanglement within cavity optomechanical architectures. However, research on nonreciprocal entanglement specifically in spinning cavity-magnon systems remains limited. Given its importance for developing robust quantum networks, devices, and circuits, there is a compelling need to explore such systems further. In this work, we propose a scheme that simultaneously leverages both the Kerr effect and the Sagnac effect to generate and control nonreciprocal magnon-magnon entanglement in a WGM cavity-magnon system.

Here, we propose a scheme to generate nonreciprocal magnon-magnon entanglement in a spinning WGM cavity-magnon system with two YIG spheres. The Kittel modes in both YIG spheres are strongly coupled to the cavity photons via magnetic dipole interactions, where the magnon Kerr effect is also taken into account. In the strong coupling regime, photons mediate entanglement between the two magnon modes, and the Sagnac effect and the Kerr effect can be harnessed to nonreciprocally enhance magnon-magnon entanglement. Furthermore, the resulting nonreciprocal entanglement exhibits robustness against bath temperature variations. Using experimentally achievable parameters, we show that the entanglement persists at bath temperatures up to approximately 100 mK. Finally, we anticipate that this work will provide new insights for the development of nonreciprocal quantum devices.

The paper is organized as follows. In Section \ref{SecII}, we introduce the system model and derive the Hamiltonian. Section \ref{SecIII} presents the effective Hamiltonian and its steady-state solutions. Section \ref{SecIV} discusses the methods for measuring entanglement and explores the effects of magnon detuning, cavity detuning, coupling strengths, and dissipation rates on magnon-magnon entanglement. Finally, Section \ref{SecV} concludes the paper with a summary of our findings and their implications for future research and applications.

\section{The model and Hamiltonian }
\label{SecII}
\begin{figure}
\centering
\includegraphics[width=0.45\textwidth]{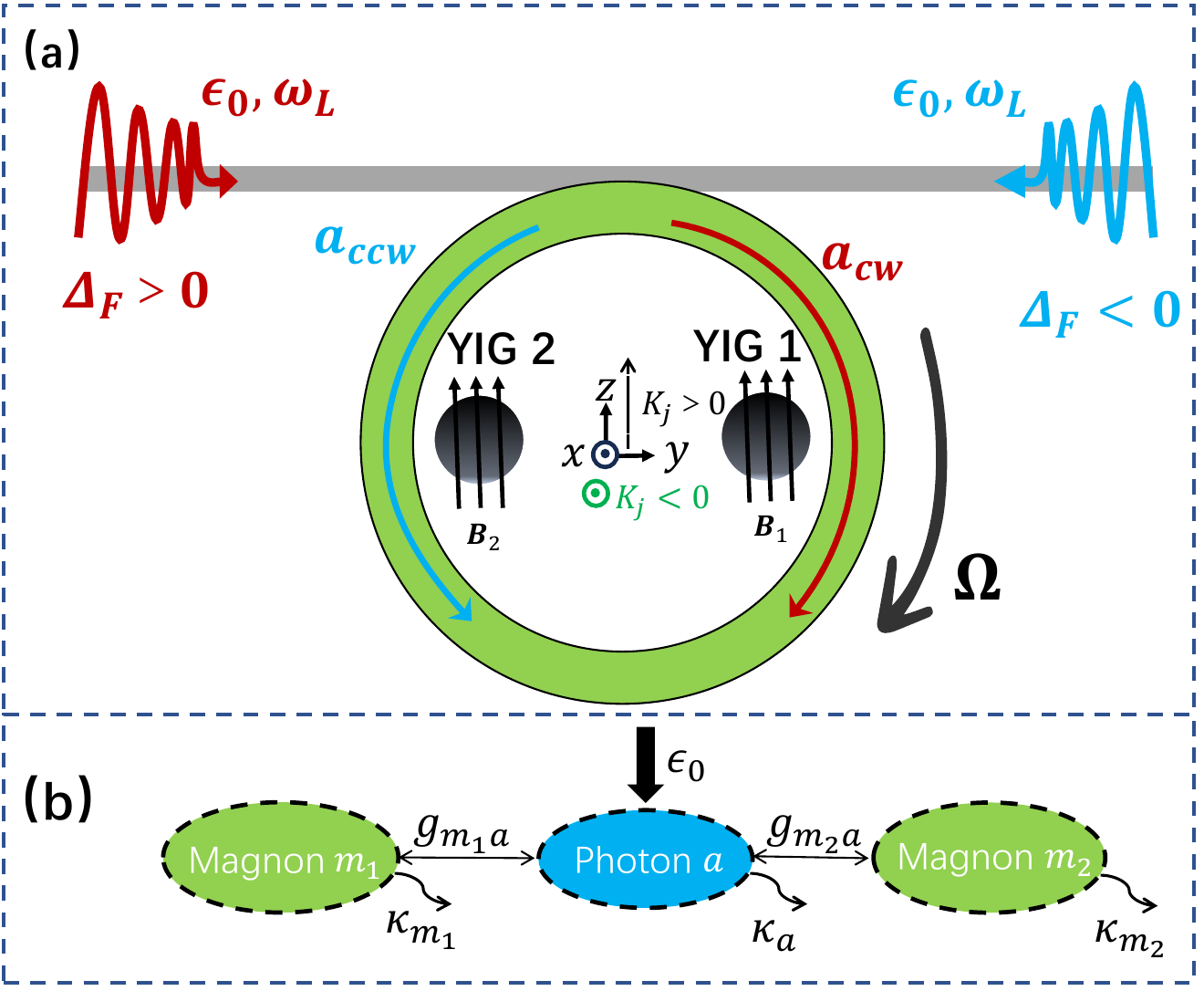}
\caption{ (a) Schematic of the spinning WGM cavity-magnon system. $K_{j}$ is the magnon Kerr coefficient, which can be adjusted by the direction of the magnetic field $B_{j}$. When the magnetic field is aligned along the crystal axis, $K_{j} > 0$; otherwise, $K_{j} < 0$. For the spinning cavity, when the driving field is clockwise (counterclockwise), a positive (negative) frequency shift $\Delta_{F} > 0$ $(< 0)$ is generated via the Sagnac effect. 
(b) The coupling configuration. Kerr magnons with decay rates $\kappa_{m_{j}}$ are coupled to photons in a spinning cavity with decay rate $\kappa_{a}$. The corresponding coupling strength is $g_{m_{j}a}$.}
\label{fig_diagram}
\end{figure}

We consider a hybrid spinning WGM cavity-magnon system, which consists of a spinning resonator with angular velocity $\Omega$ and two YIG spheres. The photon is coupled to two Kerr magnons in the Kittel mode of the \textmu m-YIG sphere [see Fig.~\ref{fig_diagram}(a)]. The Hamiltonian of this proposed system in the rotating frame with respect to the frequency of the driving field $\omega_{L}$ can be written as ($\hbar$ = 1 throughout this work)
\begin{equation}\label{eq_01}
\hat{H}=\hat{H}_{\rm SCM}+\sum_{j=1}^{2}K_{j}(\hat{m}_{j}^{\dag}\hat{m}_{j})^2+i\epsilon_0(\hat{a}^{\dag}-\hat{a}),
\end{equation}
where
\begin{equation}\label{eq_02}
\begin{split}
 \hat{H}_{\rm SCM}=&(\Delta_{a}-\Delta_{F})\hat{a}^{\dag}\hat{a}+\sum_{j=1}^{2}\Delta_{m_{j}}\hat{m}_{j}^{\dag}\hat{m}_{j}\\
 &+\sum_{j=1}^{2}g_{m_{j}a}(\hat{a}^{\dag}\hat{m}_{j}+\hat{a}\hat{m}_{j}^{\dag}),
 \end{split}
\end{equation}
where $\Delta_{a} = \omega_{a} - \omega_{L}$ is the frequency detuning between the cavity mode and the driving field, and $\omega_a$ is the resonance frequency of the non-spinning cavity. $\Delta_{m_{j}} = \omega_{m_{j}} - \omega_{L}$ is the frequency detuning between the Kittel mode and the driving field, and $\omega_{m_j} = \gamma B_j$ is the resonance frequency of the Kittel mode, determined by the gyromagnetic ratio $\gamma$ and the external bias magnetic field $B_j$. The operators $\hat{a}$ $(\hat{a}^{\dag})$ and $\hat{m}_j$ $(\hat{m}_j^{\dag})$ are the annihilation (creation) operators for the spinning cavity and the Kittel mode of the YIG sphere. $g_{m_j a}$ is the coupling coefficient between the $j$-th YIG sphere's Kittel mode and the spinning WGM cavity via the magnetic dipole interaction [see Fig.~\ref{fig_diagram}(b)]. Experiments have shown that the strong magnon-photon coupling strength is greater than the dissipation rates of the cavity mode and the Kittel mode \cite{PhysRevLett.111.127003,PhysRevLett.113.083603,PhysRevLett.113.156401}. The parameter $\Delta_F$ represents the Sagnac-Fizeau shift \cite{Malykin:2000} of the cavity resonance frequency caused by the light circulation inside the spinning cavity
\begin{equation}\label{eq_03}
\Delta_{F} = \pm \frac{\Omega n r \omega_{a}}{c} \left( 1 - \frac{1}{n^2} - \frac{\lambda}{n} \frac{dn}{d\lambda} \right),
\end{equation}
where $n$ is the refractive index, $r$ is the radius of the resonator, and $\lambda(c)$ is the wavelength (velocity) of light in vacuum. The dispersion term $dn/d\lambda$ in Eq.~\eqref{eq_03} represents the relativistic origin of the Sagnac effect, which is very small \cite{Maayani2018} ($\sim$ 0.01) and can therefore be neglected. The sign `` $+$ '' (`` $-$ '') in Eq.~\eqref{eq_03} corresponds to the clockwise (counterclockwise) driving field, where the direction of the cavity spinning is assumed to be along the clockwise direction. This implies that the clockwise (counterclockwise) driving field corresponds to $\Delta_F > 0$ ($\Delta_F < 0$).

The second term related to $K_j$ in Eq.~\eqref{eq_01} describes the Kerr nonlinearity of the magnon in the $j$-th YIG sphere Kittel mode, which is caused by the magnetocrystallographic anisotropy. The Kerr coefficient $K_j$ is inversely proportional to the volume of the YIG sphere \cite{zhang2019theory} and can be adjusted to be positive or negative by changing the direction of the magnetic field. Specifically, when the magnetic field is aligned along the crystallographic axis [100] ([110]), we have $K_j > 0$ ($< 0$) \cite{PhysRevLett.120.057202}. Experimental results show that when the diameter of the YIG sphere is reduced from 1 mm to 100 \textmu m, $K_j$ can be tuned between 0.05 and 100 \text{nHz} \cite{PhysRevB.94.224410}. The last term in Eq.~\eqref{eq_01} is the Hamiltonian of the driving field acting on the spinning cavity, where $\epsilon_0 = \sqrt{2 \kappa_a P/\omega_{L}}$ is the Rabi frequency, and $P$ is the power. 

\section{The Steady state solution And Effective Hamiltonian }
\label{SecIII}
\subsection{ Steady state solution}
The dynamics of the considered dissipative system can be described by the quantum Langevin equation $\dot{\hat{\alpha}} = i[H, \hat{\alpha}] - \kappa_{\alpha} \hat{\alpha} + \sqrt{2 \kappa_{\alpha}} \hat{\alpha}_{\text{in}}$, where $\hat{\alpha}$ represents the system operator, $\kappa_{\alpha}$ is the decay rate, and $\hat{\alpha}_{\text{in}}$ is the input noise operator.
\begin{equation}\label{eq_05}
\begin{split}
\dot{\hat{a}} =& -\left[\kappa_{a}+i(\Delta_{a}-\Delta_{F})\right]\hat{a} - \sum_{j=1}^{2}i g_{m_{j}a}\hat{m}_{j} + \epsilon_0 \\
&+ \sqrt{2\kappa_{a}}\hat{a}_{\text{in}}, \\
 \dot{\hat{m}}_{1} =& -(\kappa_{m_{1}} + i\Delta_{m_{1}})\hat{m}_{1} - i g_{m_{1}a}\hat{a} - 2i K_{1}\hat{m}_{1}^{\dagger} \hat{m}_{1} \hat{m}_{1} \\
 &+ \sqrt{2\kappa_{m_{1}}}\hat{m}_{1\text{in}}, \\
 \dot{\hat{m}}_{2}= & -(\kappa_{m_{2}} + i\Delta_{m_{2}})\hat{m}_{2} - i g_{m_{2}a}\hat{a} - 2i K_{2}\hat{m}_{2}^{\dagger} \hat{m}_{2} \hat{m}_{2}\\
 &+ \sqrt{2\kappa_{m_{2}}}\hat{m}_{2\text{in}},
\end{split}
\end{equation}
where $\hat{\sigma}_{\text{in}}$ ($\hat{\sigma} = \hat{a}, \hat{m}_j$) represents the vacuum input noise operators for the spinning cavity and the Kittel modes, respectively. The mean values of all these operators are zero, i.e., $\langle \hat{\sigma}_{\text{in}} \rangle = 0$. The correlation functions of these operators within the Markovian approximation satisfy
\begin{equation}\label{eq_06}
\begin{split}
\langle \hat{\sigma}_{in}^{\dag}(t^{\prime})\hat{\sigma}_{in}(t)\rangle=&N_{\sigma}\delta(t-t^{\prime}),\\
\langle \hat{\sigma}_{in}(t)\hat{\sigma}_{in}^{\dag}(t^{\prime})\rangle=&(N_{\sigma}+1)\delta(t-t^{\prime}),
 \end{split}
\end{equation}
where $N_{\sigma} = \left[\exp\left(\frac{\hbar \omega_{\sigma}}{k_B \text{T}}\right) - 1\right]^{-1}$ represents the average thermal excitation number for the mode $\sigma$, where $k_B$ is the Boltzmann constant and \text{T} is the bath temperature.

By rewriting each operator $\hat{\sigma}$ in Eq.~\eqref{eq_05} as the sum of its expectation value $\sigma_{s}$ and its fluctuation $\delta\hat{\sigma}$, i.e., $\hat{\sigma} = \sigma_{s} + \delta\hat{\sigma}$, the set of equations related to the operator expectation can be expressed as follows
\begin{equation}\label{eq_07}
\begin{split}
\dot{a}_{s} &= -\left[\kappa_{a}+i(\Delta_{a}-\Delta_{F})\right]a_{s}-\sum_{j=1}^{2}ig_{m_{j}a}m_{j,s}+\epsilon_0,\\
\dot{m}_{1,s} &= -\left[\kappa_{m_{1}}+i(\Delta_{m_{1}}+\Delta_{K_{1}})\right]m_{1,s}-ig_{m_{1}a}a_{s},\\
\dot{m}_{2,s} &= -\left[\kappa_{m_{2}}+i(\Delta_{m_{2}}+\Delta_{K_{2}})\right]m_{2,s}-ig_{m_{2}a}a_{s},
 \end{split}
\end{equation}
where $\Delta_{K_{j}} = 2 K_{j} |m_{j,s}|^2$ is the frequency shift caused by the Kerr effect. Considering the long-time evolution of the system, a steady state is attained with $\dot{\sigma}_{s} = 0$. Thus, Eq.~\eqref{eq_07} reduces to
\begin{equation}\label{eq_08}
\begin{split}
& 0 = \left[\kappa_{a}+i(\Delta_{a}-\Delta_{F})\right]a_{s} 
+ \sum_{j=1}^{2}ig_{m_{j}a}m_{j,s} - \epsilon_0, \\
& 0 =\left[\kappa_{m_{1}}+i(\Delta_{m_{1}}+\Delta_{K_{1}})\right]m_{1,s} 
+ ig_{m_{1}a}a_{s}, \\
& 0 = \left[\kappa_{m_{2}}+i(\Delta_{m_{2}}+\Delta_{K_{2}})\right]m_{2,s} 
+ ig_{m_{2}a}a_{s}.
 \end{split}
\end{equation}
By solving these equations, we obtain the steady-state solutions of the system
\begin{equation}\label{09}
\begin{split}
a_{s}=&\frac{\epsilon_0-\sum_{j=1}^{2}ig_{m_{j}a}m_{j,s}}{\kappa_{a}+i(\Delta_{a}-\Delta_{F})},\\
m_{1,s}=&-\frac{ig_{m_{1}a}a_{s}}{\kappa_{m_{1}}+i(\Delta_{m_{1}}+\Delta_{K_{1}})},\\
m_{2,s}=&-\frac{ig_{m_{2}a}a_{s}}{\kappa_{m_{2}}+i(\Delta_{m_{2}}+\Delta_{K_{2}})}.
 \end{split}
\end{equation}
Since \(\Delta_{F} > 0\) (\(\Delta_{F} < 0\)) depends on the direction of the clockwise (counterclockwise) driving field, the average photon number $|a_{s}|^{2}$ has different values for opposite driving directions. This indicates that $|a_{s}|^{2}$ in the spinning cavity exhibits nonreciprocal behavior. Such nonreciprocity arises from the direct coupling between the cavity and the Kittel mode of the YIG sphere, resulting in nonreciprocal average magnon numbers $|m_{j,s}|^{2}$. Notably, even for a non-spinning cavity ($\Delta_{F}$ = 0), the magnon Kerr effect induces this nonreciprocal behavior. The sign of the magnon Kerr coefficient $K_{j}$ ($K_{j} > 0$ or $K_{j} < 0$) depends on the magnetic field direction, leading to the corresponding sign of $\Delta_{K_{j}}$ ($\Delta_{K_{j}}>0$ or $\Delta_{K_{j}}<0$). This yields nonreciprocal average magnon numbers [see the second equation and the third equation in Eq.~\eqref{09}], and also induces nonreciprocity in the average photon number.

\subsection{The effective Hamiltonian}
Each operator $(\hat{\sigma})$ in Eq.~\eqref{eq_07} is rewritten as the sum of its expectation value $(\sigma_{s})$ and fluctuation $(\delta\hat{\sigma})$. The set of equations related to the operator fluctuation can be expressed as follows
\begin{equation}\label{14}
\begin{split}
\delta\dot{\hat{a}}=&-\left[\kappa_{a}+i(\Delta_{a}-\Delta_{F}) \right]\delta\hat{a}-\sum_{j=1}^{2}ig_{m_{j}a}\delta\hat{m}_{{j}}\\
&+\sqrt{2\kappa_{a}}\hat{a}_{in},\\
\delta\dot{\hat{m}}_{1}=&-\left[\kappa_{m_{1}}+i(\Delta_{m_{1}}+2\Delta_{K_{1}})\right]\delta\hat{m}_{{1}}-ig_{m_{1}a}\delta\hat{a}\\
&-2iK_{1}(m^{*}_{1,s}\delta\hat{m}^{2}_{1}+2m_{1,s}\delta\hat{m}_{1}^{\dag}\delta\hat{m}_{1}+\delta\hat{m}_{1}^{\dag}\delta\hat{m}^{2}_{1})\\
&+\sqrt{2\kappa_{m_{1}}}\hat{m}_{1,in}-2iK_{1}m^{2}_{1,s}\delta\hat{m}_{1}^{\dag},\\
\delta\dot{\hat{m}}_{2}=&-\left[\kappa_{m_{2}}+i(\Delta_{m_{2}}+2\Delta_{K_{2}})\right]\delta\hat{m}_{2}-ig_{m_{2}a}\delta\hat{a}\\
&-2iK_{2}(m^{*}_{2,s}\delta\hat{m}^{2}_2+2m_{2,s}\delta\hat{m}_{2}^{\dag}\delta\hat{m}_{2}+\delta\hat{m}_{2}^{\dag}\delta\hat{m}^{2}_{2})\\
&+\sqrt{2\kappa_{m_{2}}}\hat{m}_{2,in}-2iK_{2}m^{2}_{2,s}\delta\hat{m}_{2}^{\dag}.\\
\end{split}
\end{equation}
 For simplicity, we assume that $m_{j,s}$ is real. According to Eq.~\eqref{09}, this can be achieved by selecting an appropriate driving field phase. Under the strong driving field, the condition $ |a_{s}| \gg 1 $ can be well satisfied. Due to the beam-plitter interaction between the photon and magnon $(\hat{a}^{\dag}\hat{m}_{j} + \hat{a}\hat{m}_{j}^{\dag}) $, we also have $|m_{j,s}| \gg 1$. The higher-order fluctuation terms in Eq.~\eqref{14} are negligible, because their contributions to the system are much smaller than the linear terms. This leads to the following results
\begin{equation}\label{15}
\begin{split}
\delta\dot{\hat{a}}=&-\left[\kappa_{a}+i(\Delta_{a}-\Delta_{F}) \right]\delta\hat{a}-\sum_{j=1}^{2}ig_{m_{j}a}\delta\hat{m}_{{j}}\\
&+\sqrt{2\kappa_{a}}\hat{a}_{in},\\
\delta\dot{\hat{m}}_{1}=&-\left[\kappa_{m_{1}}+i(\Delta_{m_{1}}+2\Delta_{K_{1}})\right]\delta\hat{m}_{1}-ig_{m_{1}a}\delta\hat{a}\\
&+\sqrt{2\kappa_{m_{1}}}\hat{m}_{1,in}-i\Delta_{K_{1}}\delta\hat{m}_{1}^{\dag},\\
\delta\dot{\hat{m}}_{2}=&-\left[\kappa_{m_{2}}+i(\Delta_{m_{2}}+2\Delta_{K_{2}})\right]\delta\hat{m}_{2}-ig_{m_{2}a}\delta\hat{a}\\
&+\sqrt{2\kappa_{m_{2}}}\hat{m}_{2,in}-i\Delta_{K_{2}}\delta\hat{m}_{2}^{\dag}.
\end{split}
\end{equation}

By rewriting the above equations as $\dot{\hat{\sigma}} = i \left[\hat{H}_{\text{eff}}, \hat{\sigma} \right] - \kappa_{\sigma} \hat{\sigma} + \sqrt{2\kappa_{\sigma}} \hat{\sigma}_{\text{in}}$, the effective Hamiltonian for the linearized system can be given by
\begin{equation}\label{16}
\begin{split}
\hat{H}_{\text{eff}}=&(\Delta_{a}-\Delta_{F})\delta\hat{a}^{\dag}\delta\hat{a}+\sum_{j=1}^{2}(\Delta_{m_{j}}+2\Delta_{K_{j}})\delta\hat{m}_{j}^{\dag}\delta\hat{m}_{j}\\
&+\sum_{j=1}^{2}g_{m_{j}a}(\delta\hat{a}^{\dag}\delta\hat{m}_{j}+\delta\hat{a}\delta\hat{m}_{j}^{\dag})\\
&+\sum_{j=1}^{2}\frac{\Delta_{K_{j}}}{2}(\delta\hat{m}_{j}^{\dag}\delta\hat{m}_{j}^{\dag}+\delta\hat{m}_{j}\delta\hat{m}_{j}).
\end{split}
\end{equation}
Note that the two-magnon effect (i.e., $\delta\hat{m}_{j}^{\dag}\delta\hat{m}_{j}^{\dag}+\delta\hat{m}_{j}\delta\hat{m}_{j}$) arises from the magnon Kerr nonlinearity under the presence of a strong driving field, and it can be tuned by changing the Rabi frequency of the driving field.

\section{The Nonreciprocal magnon-magnon entanglement}
\label{SecIV}
\subsection{Entanglement metric}
\label{SecIVA}
By further defining fluctuation quadratures of the Kittle and the cavity modes as 
\begin{equation}
\hat{X}_{\sigma}=(\delta\hat{\sigma}^{\dag}+\delta\hat{\sigma})/\sqrt{2},
\hat{Y}_{\sigma}=i(\delta\hat{\sigma}^{\dag}-\delta\hat{\sigma})/\sqrt{2},
\end{equation} 
and the associated input noise operators as 
\begin{equation}
\hat{X}_{\sigma}^{in}=(\delta\hat{\sigma}_{in}^{\dag}+\delta\hat{\sigma}_{in})/\sqrt{2},
\hat{Y}_{\sigma}^{in}=i(\delta\hat{\sigma}^{\dag}_{in}-\delta\hat{\sigma}_{in})/\sqrt{2},
\end{equation}
the dynamics in Eq.~\eqref{16} can be equivalently written as
the matrix form $\dot{\hat{u}}(t) = \hat{A} \hat{u}(t) + \hat{f}(t)$, where $\hat{u}^{T}(t)$ = ($\hat{X}_{a}$, $\hat{Y}_{a}$, $\hat{X}_{m_{1}}$, $\hat{Y}_{m_{1}}$, $\hat{X}_{m_{2}}$, $\hat{Y}_{m_{2}}$) is the vector operator of the system, $\hat{f}(t)$ is the vector operator of the input noise, and
\begin{equation}
\scalebox{0.65}{%
$\hat{A} = \begin{pmatrix}
 -\kappa_a &\Delta_a -\Delta_F & 0 & g_{m_{1}a} & 0& g_{m_{2}a} \\
 -\Delta_a +\Delta_F & -\kappa_a & -g_{m_{1}a}&0 & -g_{m_{2}a} &0 \\
 0& g_{m_{1}a} & -\kappa_{m_{1}} &\Delta_{m_{1}} +\Delta_{K_{1}} &0 &0 \\
 -g_{m_{1}a} &0 & -\Delta_{m_{1}}-3\Delta_{K_{1}} & -\kappa_{m_{1}} & 0 & 0 \\
 0 & g_{m_{2}a} & 0 & 0 & -\kappa_{m_{2}} & \Delta_{m_{2}}+\Delta_{K_{2}} \\
 -g_{m_{2}a} & 0 & 0 & 0 & -\Delta_{m_{2}} - 3\Delta_{K_{2}} & -\kappa_{m_{2}}
\end{pmatrix}$},
\end{equation}
is the drift matrix. According to the Routh-Hurwitz criteria \cite{PhysRevA.35.5288}, the system is stable only when all the eigenvalues of $\hat{A}$ have negative real parts.

Due to the input quantum noise being zero-mean quantum Gaussian noise, the fluctuating quantum steady state is a zero-mean continuous-variable Gaussian state, characterized by the 6 × 6 correlation matrix $\hat{V}_{ij}^{(6)} = \langle u_{i}(\infty) u_{j}(\infty) + u_{j}(\infty) u_{i}(\infty) \rangle$ $(i,j = 1,2,...,6)$. The matrix $\hat{V}$ can be obtained by directly solving the Lyapunov equation
\begin{equation}\label{19}
\hat{A}\hat{V}+\hat{V}\hat{A}^{\dag}+\hat{D}=0,
\end{equation}
where the diffusion matrix $\hat{D}= \text{diag}[\kappa_{a}(2N_{a}+1)$, $\kappa_{a}(2N_{a}+1)$, $\kappa_{m_{1}}(2N_{m_{1}}+1)$, $\kappa_{m_{1}}(2N_{m_{1}}+1)$, $\kappa_{m_{2}}(2N_{m_{2}}+1)$, $\kappa_{m_{2}}(2N_{m_{2}}+1)]$ is defined by $\hat{D}_{ij}\delta(t-t^{\prime})=\langle v_{i}(t)v_{j}(t^{\prime})+v_{j}(t^{\prime})v_{i}(t) \rangle /2$. When the matrix $\hat{V}$ is obtained, the interest of any bipartite entanglement in the proposed system can be studied through the logarithmic negativity (The indices $\mu$ and $\nu$ denote the two subsystems of the hybrid cavity-magnon system, respectively)
\begin{equation}\label{20}
E_{N}^{(\mu\nu)}\equiv \text{max}\left\{0,-\ln\left[2\eta_{-}^{(\mu\nu)}\right]\right\},
\end{equation}
where $\eta_{-}^{(\mu\nu)}=2^{-1/2}\left\{\sum^{(\mu\nu)}-\left[\sum^{(\mu\nu)2}-4\text{det}V_{4}^{(\mu\nu)}\right]^{1/2} \right\}^{1/2}$, $\sum^{(\mu\nu)}=\text{det}V_2^{(\mu\mu)}+\text{det}V_2^{(\nu\nu)}-2\text{det}V_2^{(\mu\nu)}$, and $V_4^{(\mu\nu)} = \begin{bmatrix}
V_2^{(\mu\mu)} &V_2^{(\mu\nu)} \\
V_2^{(\nu\mu)} &V_2^{(\nu\nu)}
\end{bmatrix}$ is the 4×4 block form of the correlation matrix is associated with two modes of interest. $V_2^{(\mu\mu)}$, $V_2^{(\nu\nu)}$, $V_2^{(\mu\nu)}$ and $V_2^{(\nu\mu)}$ are the 2×2 blocks matrices of $V_4^{(\mu\nu)}$. The positive logarithmic negativity [$E_N^{(\mu\nu)} > 0$] indicates the presence of bipartite entanglement between the two modes of interest in the system under consideration.

\subsection{Exploring nonreciprocal magnon-magnon entanglement}
\begin{figure}
\centering
\includegraphics[width=0.45\textwidth]{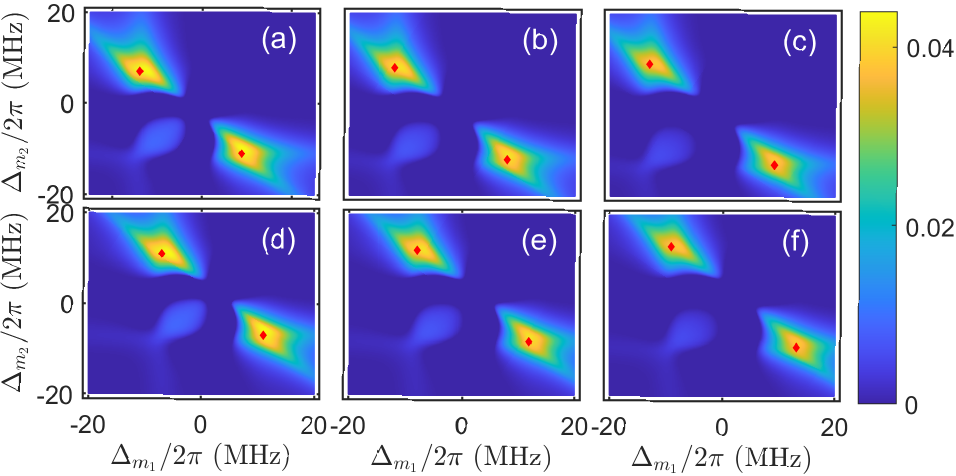}
\caption{ The relationship between the magnon-magnon entanglement $E_{N}^{m_{1}m_{2}}$ and the normalized magnon frequency detuning $\Delta_{m_{j}}/2\pi$ with 
(a) $\Delta_{K_{1}}=\Delta_{K_{2}} > 0$, $\Delta_{F} > 0$; 
(b) $\Delta_{K_{1}}=\Delta_{K_{2}} > 0$, $\Delta_{F} = 0$; 
(c) $\Delta_{K_{1}}=\Delta_{K_{2}} > 0$, $\Delta_{F} < 0$; 
(d) $\Delta_{K_{1}}=\Delta_{K_{2}} < 0$, $\Delta_{F} > 0$; 
(e) $\Delta_{K_{1}}=\Delta_{K_{2}} < 0$, $\Delta_{F} = 0$; 
(f) $\Delta_{K_{1}}=\Delta_{K_{2}} < 0$, $\Delta_{F} < 0$.}
\label{fig1}
\end{figure}

Fig.~\ref{fig1} describes the relationship between the magnon-magnon entanglement $E_{N}^{m_{1}m_{2}}$ and the normalized effective frequency detuning of the two YIG sphere's Kittel modes $\Delta_{m_{j}}/2\pi$. Using the feasible parameters: $ \Delta_{F} /2\pi = \pm 1$ and $0\text{ MHz}$, $\Delta_{K_{1}} /2\pi = \Delta_{K_{2}} /2\pi = \pm 1 \text{ MHz}$, $\omega_{a}/2\pi = 10 \text{ GHz}$, $\kappa_a/2\pi = \kappa_{m_{1}}/2\pi = \kappa_{m_{2}}/2\pi = 1 \text{ MHz}$, $g_{m_{1}a}/2\pi = g_{m_{2}a}/2\pi = 2 \text{ MHz}$, $\Delta_{a} = 10 \text{ MHz}$, $\text{T} = 10 \text{ mK}$. These parameters can ensure the system is stable according to the Routh-Hurwitz criterion.

It can be observed that the magnon-magnon entanglement can be adjusted by changing the magnon frequency detunings $\Delta_{m_{1}}$ and $\Delta_{m_{2}}$. Meanwhile, we also notice that the magnon-magnon entanglement shows nonreciprocal responses to changes in the frequency detunings $\Delta_{K_{j}}$ and $\Delta_{F}$ (i.e., Kerr and Sagnac effects), indicating that both effects can lead to nonreciprocal magnon-magnon entanglement. 
Comparing Figs.~\ref{fig1}(a) and (c), by changing the direction of the driving field (i.e., changing $\Delta_{F}$), the magnon-magnon entanglement distribution changes with the magnon effective frequency detuning. This suggests that when the Kerr effect is fixed, the nonreciprocal magnon-magnon entanglement is caused by the Sagnac effect. Similarly, in Figs.~\ref{fig1}(d) and (f), we can observe the same trend. 
By comparing Figs.~\ref{fig1}(a) and (d), we see that by adjusting the magnetic field direction (i.e., changing $\Delta_{K_{j}}$), the magnon-magnon entanglement distribution also changes with the magnon effective frequency detuning. These phenomena suggest that when the Sagnac effect is fixed, the nonreciprocal magnon-magnon entanglement is induced by the magnon Kerr effect. Similarly, in Figs.~\ref{fig1}(c) and (f), we can also observe the same trend.

\begin{figure}
\centering
\includegraphics[width=0.45\textwidth]{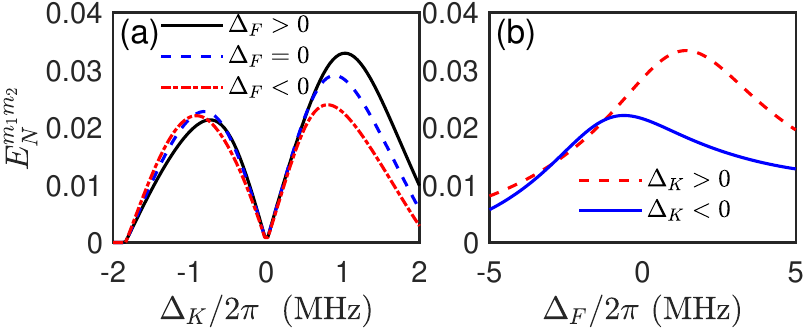}
\caption{ The impact of the normalized Kerr frequency shift $\Delta_{K}/2\pi$ and Sagnac frequency shift $\Delta_{F}/2\pi$ on the magnon-magnon entanglement $E_{N}^{m_{1}m_{2}}$ is shown with (a) $\Delta_{F} / 2\pi = \pm 1$ and $0 \text{ MHz}$; (b) $\Delta_{K} / 2\pi = \pm 1 \text{ MHz}$, where $\Delta_{m_{1}} / 2\pi = -10 \text{ MHz}$, $\Delta_{m_{2}} / 2\pi = 10 \text{ MHz}$, and other parameters are the same as those in Fig.~\ref{fig1}.
}
\label{fig2}
\end{figure}
As shown in Fig.~\ref{fig1}, we will explain the entanglement mechanism (for details, see the appendix \ref{app:squeezing}) using the feasible parameters. In Fig.~\ref{fig1}(a), the feasible parameters are $\Delta_{K_1}/2\pi = \Delta_{K_2}/2\pi =1 \text{ MHz}$, $\Delta_{F} /2\pi= 1 \text{ MHz}$ and $\Delta_{a}/2\pi = 10 \text{ MHz}$. Based on our previous derivation, the optimal entanglement frequency detuning condition should satisfy: 
$\Delta_{a} - \Delta_{F} = \pm (\Delta_{m_{1}} + 2\Delta_{K_{1}})/\cosh(2r_{1}) = \mp (\Delta_{m_{2}} + 2\Delta_{K_{2}})/\cosh(2r_{2})$. The calculated frequency detuning $\Delta_{m_{1}}/2\pi = 7.06$ $(-11.06) \text{ MHz}$, $\Delta_{m_{2}}/2\pi = -11.06$ $(7.06) \text{ MHz}$. Therefore, at the two positions that satisfy this condition, the entanglement is the largest. From Fig.~\ref{fig1}(a), we can observe that the numerical simulation results (the red diamonds represent the numerical simulation results) agree well with the picture. Next, we use the optimal entanglement condition to calculate Fig.~\ref{fig1}(c). The used parameters are set as follows: $\Delta_{K_1} /2\pi= \Delta_{K_2} /2\pi= 1 \text{ MHz}$, $\Delta_{F}/2\pi = -1 \text{ MHz}$, $\Delta_{a}/2\pi = 10 \text{ MHz}$. Substituting the optimal entanglement condition, the calculated frequency detunings are $\Delta_{m_{1}}/2\pi = 9.05$ $(-13.05) \text{ MHz}$ and $\Delta_{m_{2}}/2\pi = -13.05$ $(9.05) \text{ MHz}$. This matches the optimal entanglement position in the picture. Meanwhile, numerical simulations have also been performed under other parameter conditions, as shown in Fig.~\ref{fig1}. This verifies the validity of the optimal magnon-magnon entanglement theory obtained via the squeezing framework.

In experiments, both the magnon Kerr frequency shift and Sagnac frequency shift are tunable. Therefore, it is necessary to explore how these two factors influence magnon-magnon entanglement. Fig.~\ref{fig2} describes the relationship between magnon-magnon entanglement and the normalized Sagnac and Kerr frequency shifts. We can observe that when the Kerr effect is present, magnon-magnon entanglement is significantly enhanced. By fixing $ \Delta_{K_{1}} = \Delta_{K_{2}} = \Delta_{K}$, we observe in Fig.~\ref{fig2}(a) that as the Kerr frequency shift increases from zero in both directions, the entanglement first increases and then decreases, with the optimal entanglement corresponding to the Kerr frequency shift $ \Delta_{K}/ 2\pi \approx \pm 1 \text{ MHz}$. In Fig.~\ref{fig2}(b), we can see that as the Sagnac frequency shift increases, the entanglement first increases and then decreases, with the optimal entanglement corresponding to the Sagnac frequency shift $\Delta_{F}/ 2\pi \approx \pm 1 \text{ MHz}$. Therefore, both the Kerr effect and the Sagnac effect can enhance entanglement of the system. To achieve the higher magnon-magnon entanglement, $\Delta_{F}/2\pi = \Delta_{K}/2\pi = \pm 1 \text{ MHz}$ will be used in the paper.

\subsection{The effect of cavity field frequency detuning on the magnon-magnon entanglement}
\begin{figure}
\centering
\includegraphics[width=0.45\textwidth]{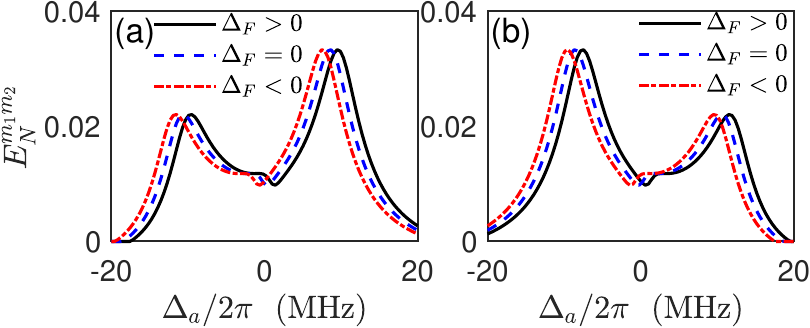}
\caption{ The effect of normalized cavity field frequency detuning $\Delta_{a}/2\pi$ on magnon-magnon entanglement $E_{N}^{m_{1}m_{2}}$.
Parameter settings: $\Delta_{m_{1}}/2\pi=-10\text{ MHz}$, $\Delta_{m_{2}}/2\pi=10\text{ MHz}$, $\Delta_{F} / 2\pi = \pm 1$ and $0 \text{ MHz}$, 
(a) $\Delta_{K_{1}}=\Delta_{K_{2}} =1 \text{ MHz}$; 
(b) $\Delta_{K_{1}}=\Delta_{K_{2}} =-1 \text{ MHz}$. Other parameters are the same as those in Fig.~\ref{fig1}.}
\label{fig3}
\end{figure}
In the previous discussion, we explored the nonreciprocal magnon-magnon entanglement induced by the Kerr effect and the Sagnac effect, and investigated the optimal entanglement positions under different magnon frequency detunings. We fix $\Delta_{m_{1}}/2\pi=-10 \text{ MHz}$, $\Delta_{m_{2}}/2\pi=10 \text{ MHz}$ to study the effect of cavity field frequency detuning on entanglement. Fig.~\ref{fig3} shows the relationship between magnon-magnon entanglement $E_{N}^{m_{1}m_{2}}$ and normalized cavity field frequency detuning $\Delta_{a}/2\pi$. From Figs. \ref{fig3}(a) and (b), we can clearly observe the nonreciprocal entanglement caused by both effects. When the direction of the driving field or the magnetic field is changed, the distribution of entanglement shifts. Additionally, as the cavity field frequency detuning $\Delta_a$ increases, the magnon-magnon entanglement reaches the maximum at two positions. As the cavity field frequency detuning continues to increase, the magnon-magnon entanglement disappears. This is because the optimal entanglement position is jointly controlled by both cavity field frequency detuning and magnon frequency detuning, and must satisfy the condition 
$\Delta_{a} - \Delta_{F} = \pm (\Delta_{m_{1}} + 2\Delta_{K_{1}})/\cosh(2r_{1}) = \mp (\Delta_{m_{2}} + 2\Delta_{K_{2}})/\cosh(2r_{2})$. Thus, when the direction of the driving field and the magnetic field is changed, the corresponding cavity field frequency detuning for optimal entanglement will shift. In particular, we find that when the Kerr effect is fixed, as the Sagnac frequency shift changes from positive to negative, the cavity field frequency detuning for optimal entanglement gradually decreases.
\begin{figure}
\centering
\includegraphics[width=0.45\textwidth]{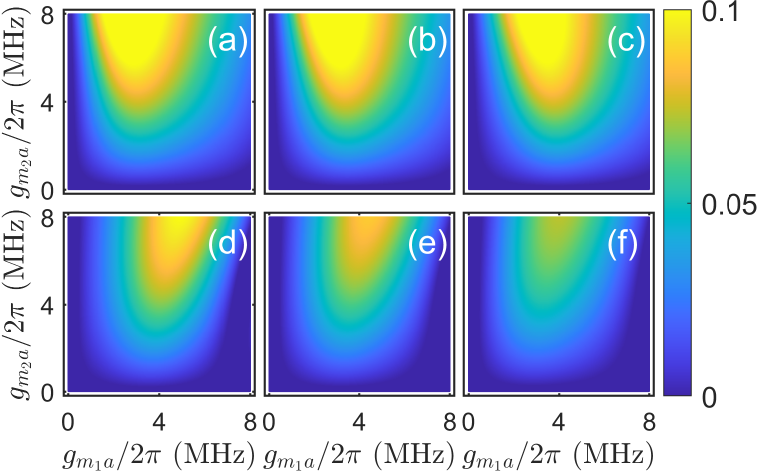}
\caption{ The effect of the normalized magnon-photon coupling strengths $g_{m_{j}a}/2\pi$ on the magnon-magnon entanglement $E_{N}^{m_{1}m_{2}}$ with 
(a) $\Delta_{K_{1}}=\Delta_{K_{2}} > 0$, $\Delta_{F} > 0$;
(b) $\Delta_{K_{1}}=\Delta_{K_{2}} > 0$, $\Delta_{F} = 0$;
(c) $\Delta_{K_{1}}=\Delta_{K_{2}} > 0$, $\Delta_{F} < 0$;
(d) $\Delta_{K_{1}}=\Delta_{K_{2}} < 0$, $\Delta_{F} > 0$;
(e) $\Delta_{K_{1}}=\Delta_{K_{2}} < 0$, $\Delta_{F} = 0$;
(f) $\Delta_{K_{1}}=\Delta_{K_{2}} < 0$, $\Delta_{F} < 0$, 
where $\Delta_{m_{1}}/2\pi = -10\text{ MHz}$, $\Delta_{m_{2}}/2\pi = 10\text{ MHz}$, $\Delta_{F} / 2\pi = \pm 1$ and $0 \text{ MHz}$, $\Delta_{K_{1}} / 2\pi = \Delta_{K_{2}} / 2\pi = \pm 1 \text{ MHz}$ and other parameters are the same as those in Fig.~\ref{fig1}.}
\label{fig4}
\end{figure}
\subsection{The effect of magnon-photon coupling strength on magnon-magnon entanglement}
In the previous exploration, we assumed that the coupling strengths of two YIG spheres and the cavity mode are the same. But what happens to the magnon-magnon entanglement when the coupling strengths differ? To the end, we plot the relationship between the magnon-magnon entanglement $E_{N}^{m_{1}m_{2}}$ and the normalized magnon-photon coupling strengths $g_{m_{j}a}/2\pi$. It is clear that the magnon-photon coupling strengths have a significant effect on the entanglement. From Fig.~\ref{fig4}, we not only observe the nonreciprocal behavior of magnon-magnon entanglement, but also find that as the coupling strengths $g_{m_{1}a}$ and $g_{m_{2}a}$ increase, the entanglement is significantly enhanced. Additionally, comparing Figs. \ref{fig4}(a) and (d), with the Sagnac effect fixed, when the Kerr coefficient is negative, the coupling strength for the optimal entanglement is bigger, while when the Kerr coefficient is positive, the coupling strength for the optimal entanglement is smaller. However, when we fix the Kerr effect and compare Figs.~\ref{fig4}(a) and (c), the entanglement corresponding to a positive Kerr coefficient is larger. Therefore, appropriately adjusting the magnon-photon coupling strength will have a significant effect on the entanglement.
\begin{figure}
\centering
\includegraphics[width=0.45\textwidth]{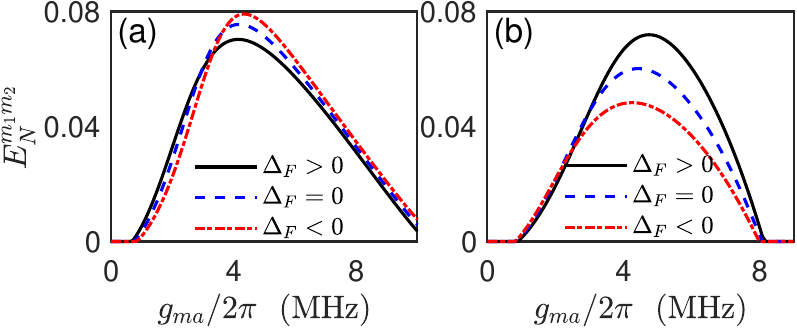}
\caption{ The effect of the normalized magnon-photon coupling strength $g_{ma}/2\pi$ on the magnon-magnon entanglement $E_{N}^{m_{1}m_{2}}$ with 
(a) $\Delta_{K_{1}} = \Delta_{K_{2}} =1 \text{ MHz}$;
(b) $\Delta_{K_{1}} = \Delta_{K_{2}}=-1 \text{ MHz}$, 
where $\Delta_{m_{1}}/2\pi = -10\text{ MHz}$, $\Delta_{m_{2}}/2\pi = 10\text{ MHz}$, $\Delta_{F} / 2\pi = \pm 1$ and $0 \text{ MHz}$, and other parameters are the same as those in Fig.~\ref{fig1}.}
\label{fig5}
\end{figure}

Fig.~\ref{fig5} further illustrates the effect of coupling strength on the nonreciprocal magnon-magnon entanglement. When $g_{m_{1}a} = g_{m_{2}a} = g_{ma}$ is fixed, as the coupling strength increases to $g_{ma}/2\pi \approx 4 \text{ MHz}$, the optimal entanglement is reached. If the coupling strength continues to increase, the entanglement decreases. This is because the initial increase in coupling strength enhances the interaction between the optical field mode and the magnon modes, thereby promoting the generation of entanglement. However, when the coupling strength becomes too large, nonlinear effects of the system and the imbalance in the coupling strength lead to a weakening of the entanglement.

\subsection{Effect of dissipation on magnon-magnon entanglement}
\begin{figure}
\centering
\includegraphics[width=0.45\textwidth]{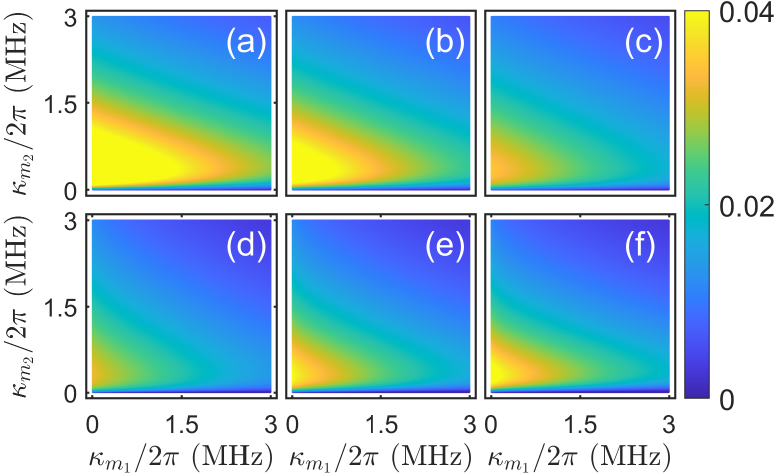}
\caption{ The effect of the normalized decay rates of the magnon $\kappa_{m_{j}}/2\pi$ on the magnon-magnon entanglement $E_{N}^{m_{1}m_{2}}$ with 
(a) $\Delta_{K_{1}}=\Delta_{K_{2}} > 0$, $\Delta_{F} > 0$; 
(b) $\Delta_{K_{1}}=\Delta_{K_{2}} > 0$, $\Delta_{F} = 0$; 
(c) $\Delta_{K_{1}}=\Delta_{K_{2}} > 0$, $\Delta_{F} < 0$; 
(d) $\Delta_{K_{1}}=\Delta_{K_{2}} < 0$, $\Delta_{F} > 0$; 
(e) $\Delta_{K_{1}}=\Delta_{K_{2}} < 0$, $\Delta_{F} = 0$; 
(f) $\Delta_{K_{1}}=\Delta_{K_{2}} < 0$, $\Delta_{F} < 0$, where $\Delta_{m_{1}}/2\pi = -10$ MHz, $\Delta_{m_{2}}/2\pi = 10$ MHz, $\Delta_{F} / 2\pi = \pm 1$ and $0 \text{ MHz}$, $\Delta_{K_{1}} / 2\pi = \Delta_{K_{2}} / 2\pi = \pm 1 \text{ MHz}$. Other parameters are the same as those in Fig.~\ref{fig1}.}
\label{fig6}
\end{figure}
Previously, we demonstrated the effects of the magnon effective frequency detuning, the cavity field frequency detuning, and the magnon-photon coupling strengths on the entanglement. In addition, the decay rates of the magnon can also be adjusted in experiments. Therefore, it's important to consider the impact of the decay rate of the magnon. 

Fig.~\ref{fig6} shows the relationship between the magnon-magnon entanglement $E_{N}^{m_{1}m_{2}}$ and the normalized decay rates $\kappa_{m_{j}}/2\pi$. When other parameters are fixed, adjusting $\kappa_{m_{j}}$ can achieve optimal entanglement. We found that entanglement gradually decreases as the decay rate increases. Entanglement is maintained at a high level only when the decay rate is low. Therefore, the decay rate has a significant effect on magnon-magnon entanglement.

\begin{figure}
\centering
\includegraphics[width=0.45\textwidth]{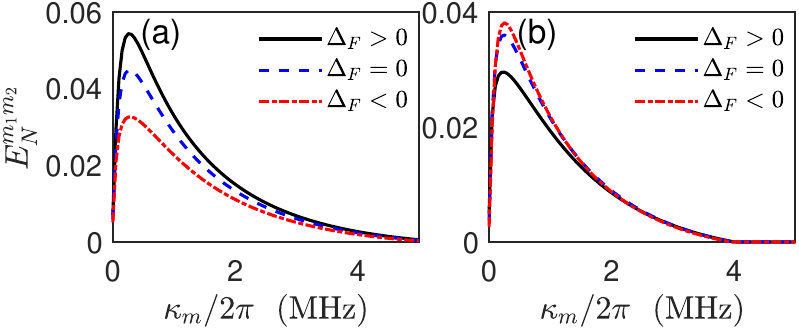}
\caption{ The effect of the normalized decay rate of the magnon $\kappa_{m}/2\pi$ on the magnon-magnon entanglement $E_{N}^{m_{1}m_{2}}$ with 
(a) $\Delta_{K_{1}} = \Delta_{K_{2}} =1 \text{ MHz}$; 
(b) $\Delta_{K_{1}} = \Delta_{K_{2}} =-1 \text{ MHz}$, where $\Delta_{m_{1}}/2\pi = -10\text{ MHz}$, $\Delta_{m_{2}}/2\pi = 10\text{ MHz}$, $\Delta_{F} / 2\pi = \pm 1$ and $0 \text{ MHz}$. Other parameters are the same as those in Fig.~\ref{fig1}.}
\label{fig7}
\end{figure}

From Fig.~\ref{fig7}, by fixing $\kappa_{m_{1}}$ = $\kappa_{m_{2}}$ = $\kappa_{m}$, we further investigate the effect of the decay rate. By comparing Figs.~\ref{fig7}(a) and (b), we observe the nonreciprocal entanglement induced by the Kerr effect and the Sagnac effect. Notably, when the Kerr coefficients are different, the decay rates corresponding to the disappearance of entanglement are different. When the Kerr coefficient is positive, the maximum decay rate of the system is higher. When the Kerr coefficient is negative, the maximum decay rate is lower. Therefore, we can increase the decay rate of the magnon nonreciprocally in the system by utilizing the Kerr effect. From Fig.~\ref{fig7}, we can conclude that a large decay rate leads to the disappearance of entanglement. This is because when the decay rate is small, the interaction between the cavity mode and the magnon mode promotes the increase of quantum entanglement. The coherence of the system is strong, which is favorable for the accumulation of entanglement. As the decay rate increases, the coupling between the system and the environment becomes more significant, leading to the loss of quantum coherence and thus reducing entanglement. Additionally, we observe that when $\Delta_{K_{j}}\Delta_F > 0$, the magnon-magnon entanglement corresponding to the same decay rate is larger. This indicates that high entanglement can be obtained by utilizing the positive cooperative effect (i.e., when the Kerr effect and Sagnac effect coefficients have the same sign). When considering the negative cooperative effect of the Kerr and Sagnac effects (i.e., when the Kerr and Sagnac effect coefficients have opposite sign), the entanglement corresponding to the same decay rate decreases.
\begin{figure}
\includegraphics[width=0.45\textwidth]{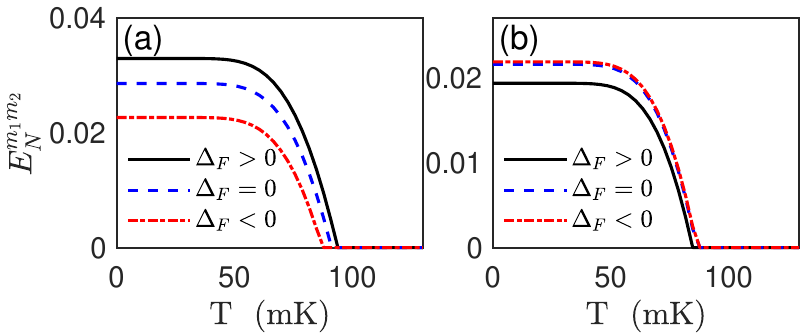}
\caption{ The effect of bath temperature \text{T} on the magnon-magnon entanglement $E_{N}^{m_{1}m_{2}}$ with
(a) $\Delta_{K_{1}} = \Delta_{K_{2}} =1\text{ MHz}$; 
(b) $\Delta_{K_{1}} = \Delta_{K_{2}} =-1\text{ MHz}$,
where $\Delta_{m_{1}}/2\pi = -10\text{ MHz}$, $\Delta_{m_{2}}/2\pi = 10\text{ MHz}$, $\Delta_{F} / 2\pi = \pm 1$ and $0 \text{ MHz}$, and other parameters are the same as those in Fig.~\ref{fig1}.}
\label{fig8}
\end{figure}
\subsection{The effect of bath temperature on magnon-magnon entanglement}
We examine the effect of bath temperature \text{T} on the magnon-magnon entanglement in our scheme. We plot the relationship between the magnon-magnon entanglement $E_{N}^{m_{1}m_{2}}$ and the bath temperature \text{T}. It is evident that when the bath temperature reaches a certain value, the magnon-magnon entanglement disappears. We define the critical temperature at which entanglement vanishes as the survival temperature. This is because, as the bath temperature increases, thermal noise becomes stronger, and thermal noise causes decoherence effects in the quantum state, which destroys the quantum coherence in the system, thereby reducing the quantum entanglement. Meanwhile, we observe that the bath temperature and dissipation effects have the same synergistic effect, that is, when $\Delta_{K_{j}}\Delta_{F} > 0$, the magnon-magnon entanglement corresponding to the same bath temperature increases, and when $\Delta_{K_{j}}\Delta_F < 0$, the magnon-magnon entanglement corresponding to the same bath temperature decreases. Finally, as shown in Fig.~\ref{fig8}(a), by fixing the Kerr effect, the survival bath temperature of the magnon entanglement can be nonreciprocally increased by the Sagnac effect [see the 
red and black lines in Fig.~\ref{fig8}(a)]. In contrast, by fixing the Sagnac effect, as shown in Figs.~\ref{fig8}(a) and (b) blue line, the Kerr effect also nonreciprocally increases the survival bath temperature of the entanglement. Considering both effects together, the survival bath temperature of the magnon-magnon entanglement nonreciprocally increases to approximately 100 mK.

\section{CONCLUSION}
\label{SecV}
In summary, we propose a new scheme to generate nonreciprocal magnon-magnon entanglement in a spinning WGM cavity-magnon system by utilizing the Kerr effect and the Sagnac effect. The two magnons and the photon are coupled through magnetic dipole interactions, and the entanglement between the two magnons is mediated by the photon. Considering the Kerr effect and the Sagnac effect, the entanglement can be nonreciprocally enhanced or suppressed. We find that by adjusting parameters such as the dissipation rate, coupling strength, and cavity field frequency detuning, optimal magnon-magnon entanglement can be achieved. Among these, the coupling strength significantly affects the entanglement, while high dissipation rates can destroy the entanglement. Finally, in our scheme, the survival bath temperature of the entanglement can reach approximately 100 \text{mK}. Our work paves the way for designing nonreciprocal devices using the Kerr effect and the Sagnac effect in hybrid spinning cavity-magnon systems.

\section*{Acknowledgments}
This work is supported by the Innovation Program for Quantum Science and Technology (No. 2023ZD0300700).

\begin{widetext}
\appendix
\renewcommand{\appendixname}{APPENDIX}
\appendix
\section{Magnon squeezing transformation and optimal resonance condition}
\label{app:squeezing}
In this appendix, we derive the optimal resonance condition for magnon-magnon entanglement by performing a unified squeezing transformation \cite{PhysRevB.105.245310} on both magnon modes. The two individual squeezing operations are combined into a single unitary transformation to simplify the derivation.

We start with the linearized effective Hamiltonian $\hat{H}_\text{eff}$ given in Eq.~\eqref{16} of the main text. The joint unitary squeezing operator for the two magnon modes is defined as
\begin{equation}
\hat{U}=\hat{U}_1\hat{U}_2
=\exp\left[\frac{r_1}{2}\left(\hat{m}_1^2-\hat{m}_1^{\dagger2}\right)\right]
\exp\left[\frac{r_2}{2}\left(\hat{m}_2^2-\hat{m}_2^{\dagger2}\right)\right],
\end{equation}
where the squeezing parameters are
\begin{equation}
r_{j}=\frac{1}{4}\ln\left(\frac{\Delta_{m_j}+\Delta_{K_j}}{\Delta_{m_j}+3\Delta_{K_j}}\right).
\end{equation}
Under the transformation $\hat{\beta}_j=\hat{U}^\dagger \hat{m}_j \hat{U}$, the original magnon operators are mapped to the squeezed bosonic operators $\hat{\beta}_j=\hat{m}_j\cosh r_j-\hat{m}_j^\dagger\sinh r_j$, with the inverse relations $\hat{m}_j=\hat{\beta}_j\cosh r_j+\hat{\beta}_j^\dagger\sinh r_j$.
We can verify the squeezed operators $\hat{\beta}_j$ obey standard bosonic commutation relations.

Substituting the inverse transformations into $\hat{H}_\text{eff}$ and neglecting non-resonant high-order terms under strong driving, we obtain the Hamiltonian in the joint squeezing frame:
\begin{equation}
\begin{aligned}
\hat{H}_\text{squeezed}
&=(\Delta_a-\Delta_F)\hat{a}^\dagger\hat{a}
+\frac{\Delta_{m_1}+2\Delta_{K_1}}{\cosh2r_1}\hat{\beta}_1^\dagger\hat{\beta}_1
+\frac{\Delta_{m_2}+2\Delta_{K_2}}{\cosh2r_2}\hat{\beta}_2^\dagger\hat{\beta}_2+g_{m_1a}\cosh r_1\left(\hat{a}^\dagger\hat{\beta}_1+\hat{a}\hat{\beta}_1^\dagger\right)
+g_{m_2a}\cosh r_2\left(\hat{a}^\dagger\hat{\beta}_2+\hat{a}\hat{\beta}_2^\dagger\right)\\
&\quad+g_{m_1a}\sinh r_1\left(\hat{a}^\dagger\hat{\beta}_1^\dagger e^{2it(\Delta_a-\Delta_F)}+\text{H.c.}\right)+g_{m_2a}\sinh r_2\left(\hat{a}^\dagger\hat{\beta}_2^\dagger e^{2it(\Delta_a-\Delta_F)}+\text{H.c.}\right).
\end{aligned}
\end{equation}

We then move to the interaction picture by removing the free evolution terms of the cavity and squeezed magnon modes. To achieve resonant magnon-magnon entanglement, the time-dependent oscillating phase factors must vanish, yielding the key optimal resonance condition:
\begin{equation}
\Delta_a-\Delta_F
=\pm\frac{\Delta_{m_1}+2\Delta_{K_1}}{\cosh2r_1}
=\mp\frac{\Delta_{m_2}+2\Delta_{K_2}}{\cosh2r_2}.
\end{equation}

Under this condition, only the static resonant interaction terms remain, giving the simplified interaction Hamiltonian
\begin{equation}
\hat{H}_\text{int}
=g_{m_1a}\cosh r_1\left(\hat{a}^\dagger\hat{\beta}_1+\hat{a}\hat{\beta}_1^\dagger\right)
+g_{m_2a}\cosh r_2\left(\hat{a}^\dagger\hat{\beta}_2+\hat{a}\hat{\beta}_2^\dagger\right).
\end{equation}
This Hamiltonian describes the photon-mediated coherent coupling between the two squeezed magnon modes, which dominates the generation and enhancement of nonreciprocal magnon-magnon entanglement.
\end{widetext}
\bibliography{ref}
\vspace{8pt}
\end{document}